\documentclass[sigconf, 10pt]{acmart}
\copyrightyear{2026}\acmYear{2026}
\setcopyright{cc}
\setcctype{by}
\acmConference[HotNets '26]{The 25th ACM Workshop on Hot Topics in Networks}{November 16-17, 2026}{Salt Lake City, UT, USA}
\acmBooktitle{The 25th ACM Workshop on Hot Topics in Networks (HotNets '26), November 16-17, 2026, Salt Lake City, UT, USA}
\acmDOI{10.1145/3847290.3859940}
\acmISBN{979-8-4007-3029-0/2026/11}

\usepackage{tikz}
\usetikzlibrary{arrows.meta,positioning,fit,calc,backgrounds,shadows.blur}
 
\definecolor{cReq}{HTML}{2C5FE0}   
\definecolor{cResp}{HTML}{0E9C86}  
\definecolor{cIO}{HTML}{7C3AED}    
\definecolor{cBG}{HTML}{D9831A}    
\definecolor{boxFill}{HTML}{EAF1FF}
\definecolor{boxEdge}{HTML}{3B6FD4}
\definecolor{sFill}{HTML}{E7F6EC}
\definecolor{sEdge}{HTML}{2F9E5E}
\definecolor{origFill}{HTML}{F1F2F4}
\definecolor{origEdge}{HTML}{9AA0A6}
\definecolor{scdnFill}{HTML}{F5F7FA}
\definecolor{panelA}{HTML}{F6F9FF}
\definecolor{panelB}{HTML}{F5FBF7}
\definecolor{inkLbl}{HTML}{2A2E37}
\definecolor{iFill}{HTML}{E9EEFA}\definecolor{pFill}{HTML}{FBEAD2}
\definecolor{pEdge}{HTML}{C9781A}\definecolor{scdnFill}{HTML}{F3F7FF}

\newif\ifcomment
\commentfalse

\ifcomment
    \newcounter{YXNumberOfComments}
    \stepcounter{YXNumberOfComments}
    \newcommand{\xym}[1]{\textcolor{orange}{\small \bf [XYM\#\arabic{YXNumberOfComments}\stepcounter{YXNumberOfComments}: #1]}}

    \newcounter{PHNumberOfComments}
    \stepcounter{PHNumberOfComments}
    \newcommand{\hpc}[1]{\textcolor{blue}{\small \bf [HPC\#\arabic{PHNumberOfComments}\stepcounter{PHNumberOfComments}: #1]}}

    \newcounter{TODONumberOfComments}
    \stepcounter{TODONumberOfComments}
    \newcommand{\todo}[1]{\textcolor{red}{\small \bf [TODO\#\arabic{TODONumberOfComments}\stepcounter{TODONumberOfComments}: #1]}}  
    
    \newcommand{\del}[1]{{{\color{red}\st{#1}}}}
\else
    \newcommand\xym[1]{}    
    \newcommand\hpc[1]{} 
    \newcommand\todo[1]{}
    \newcommand{\del}[1]{}
\fi

\newif\ifrevision
\revisionfalse
\definecolor{revColor}{HTML}{1F4FD8}
\ifrevision
    \newcommand{\rev}[1]{{\color{revColor}#1}}
\else
    \newcommand{\rev}[1]{#1}
\fi

\usepackage{booktabs}
\usepackage{subcaption}

\begin{document}

\title{Semantics Delivery Network: Rethinking Web Retrieval Infrastructure for LLM Agents}

\author{Peichun Hua}
\affiliation{%
  \institution{The Chinese University of Hong Kong, Shenzhen}
  \city{Shenzhen, Guangdong}
  \country{China}
}
\email{peichunhua@link.cuhk.edu.cn}

\author{Yunming Xiao}
\authornote{Corresponding author}
\affiliation{%
  \institution{The Chinese University of Hong Kong, Shenzhen}
  \city{Shenzhen, Guangdong}
  \country{China}
}
\email{yunmingxiao@cuhk.edu.cn}

\begin{abstract}
Large language models (LLMs) increasingly ground their answers in proprietary data and live web content through both single-shot retrieval-augmented generation (RAG) and multi-turn agentic search. Yet the web path they use was built for human readers: search services return URLs and snippets ranked by generic relevance, and content delivery networks (CDNs) cache URL-addressed objects without knowing which passage an agent needs.
LLMs instead consume short, semantically coherent passages—hereafter, \textit{chunks}—selected for downstream utility rather than similarity alone, retrieve statefully across reasoning turns, and, as uncoordinated streams, repeatedly fetch, clean, chunk, and embed the same content.

We argue that \emph{semantic chunk retrieval should become a first-class network-delivery abstraction}, and propose the \emph{Semantics Delivery Network} (SemDN): an origin-authorized, hierarchical substrate that indexes, searches, and caches web content at chunk granularity, serving clients and origins alike. 
\rev{SemDN is more than finer-grained caching: it must estimate missing or stale evidence without an explicit miss signal, let clients choose retrieval representations that origins need not see, and share state among clients with the same representation.}
Preliminary results show that page-first retrieval processes two to three orders of magnitude more content than agents consume, that agentic queries exhibit task-local chunk reuse invisible to URL caches, and that chunk delivery improves answer quality per context token.
We organize an open research agenda around shareable retrieval state, hierarchical caching and refresh under coverage risk, and deployability, including incentives, privacy, and verifiability. Our code is available at: \url{https://github.com/sarendis56/SemDN_HotNets26}.
\end{abstract}



\maketitle

\section{Introduction}

Machine-generated retrieval is becoming an important part of the web's read path. LLMs retrieve external content to ground answers, from single-shot RAG~\cite{lewis2020rag} to multi-turn \emph{agentic search} that interleaves reasoning and retrieval~\cite{jin2025searchr1,liu2026agenticr}. Increasingly, this retrieval targets the \emph{live} web: browsing agents and freshness-sensitive benchmarks make static corpora insufficient~\cite{nakano2021webgpt,wei2025browsecomp,wu2025webwalker,vu2023freshllms,lazaridou2021gap,liska2022streamingqa}, while query-time fetcher traffic is rising sharply~\cite{cloudflare,fastly,zhang2025aiwebcache}.

%

\textbf{Yet these machine clients retrieve through infrastructure designed for human readers.} Discovery APIs expose URL- and snippet-oriented results optimized for generic relevance; CDNs cache URL-addressed objects without knowing which passage an agent needs. An LLM instead consumes short semantic chunks\footnote{A \emph{chunk} broadly denotes a semantic unit; this may be a text passage~\cite{karpukhin2020dpr}, an image region~\cite{radford2021clip,faysse2024colpali} or page screenshot~\cite{wang2026pixelrag}, a table block~\cite{herzig2021tableretrieval}, an audio clip~\cite{wu2023clap}, or a video segment~\cite{luo2022clip4clip}.} rather than pages~\cite{zhang2025sage,qu2025semantic}.
It cares about downstream \emph{utility}, not similarity alone~\cite{liu2026agenticr,seper2025lu,shi2024replug,chandra2026lurerag}; retrieves statefully across related sub-queries; and belongs to an uncoordinated population that repeatedly downloads, cleans, and embeds the same content, pushing redundant work onto origins~\cite{fastly}.

\begin{figure*}[thbp]
\centering
\includegraphics[width=\textwidth]{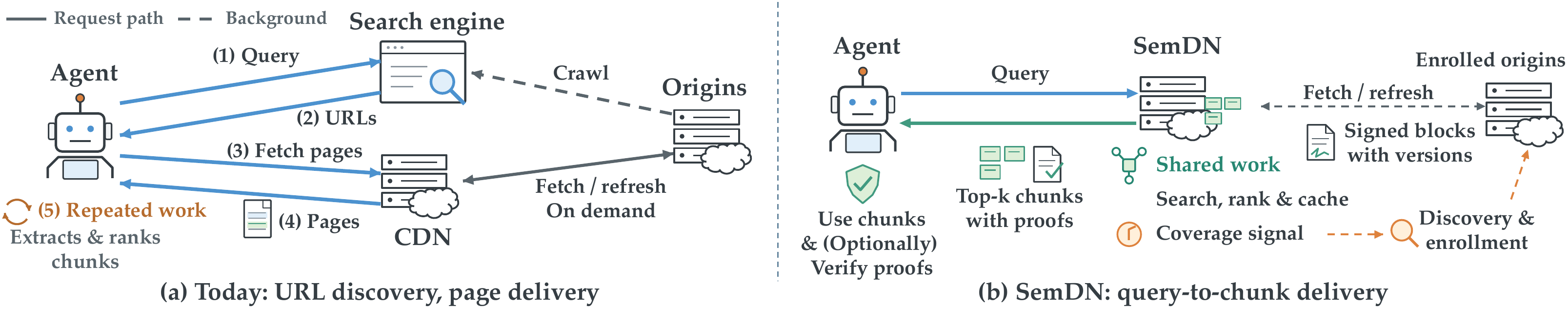}
\caption{Today's retrieval path versus SemDN. \textbf{(a)}~A search engine returns URLs; the agent fetches whole pages through a semantics-agnostic CDN and extracts chunks itself. \textbf{(b)}~SemDN answers the query with top-$k$ chunks from a shared index over enrolled origins, optionally with proofs; a coverage signal triggers discovery and enrollment.}
\label{fig:arch}
\vspace{-0.05in}
\end{figure*}

\rev{Today, this path is split across two systems (Fig.~\ref{fig:arch}a). A search engine serves the client and decides \emph{what} to read, but it returns only URLs. A CDN serves the origin with edge placement, offload, freshness management~\cite{ali2025cdn}, and even origin logic at the edge~\cite{lin2025preacher,cloudflare-worker-example}, but it never sees the query.}
We argue that \emph{semantic chunk retrieval should become a first-class network-delivery abstraction}. We propose the \emph{Semantics Delivery Network} (SemDN), \rev{a semantic delivery substrate that fuses both roles: like a search engine, it answers clients' queries; like a CDN, it delivers origin content from the edge.} It encodes, indexes, searches, and caches web content at chunk granularity, returning the passages an LLM needs rather than pages it must download and post-process.

\rev{SemDN could be built atop an existing CDN or as a separate system, but it is more than finer-grained caching.
Keying delivery on queries instead of URLs changes three things. (1) \emph{Misses:} a URL lookup misses on an exact name, whereas top-$k$ search always answers, so missing or stale evidence must be estimated. (2) \emph{Tenancy:} origins still enroll for offload and protection, as with a CDN, but each client chooses the encoder and index it retrieves with, a choice origins need not see. (3) \emph{State:} a cached page serves every client, whereas index and cache state serve only clients sharing a representation; only acquisition and normalization are shared by all. Each difference calls for careful design to attain performance and efficiency.}

Two recent results make this new paradigm increasingly practical.
First, recompute-based indexing, such as EdgeRAG and LEANN, shows that embeddings need not all be stored but can be regenerated on demand during search~\cite{wang2025leann,seemakhupt2024edgerag}. An edge can therefore keep content blocks and materialize model-specific vectors only where demand justifies them.
Second, agent query streams can be bursty and task-coherent, creating locality that an edge hierarchy may exploit for placement and reuse~\cite{lee2026agenttrace,deng2026lever}.

Prior systems separately expose retrieved chunks as a service~\cite{almaliki2026caas}, build managed indexes over tenant corpora~\cite{cloudflareaisearch}, \rev{tier retrieval across edge and cloud for one deployment~\cite{li2024eacorag},} or reduce vector-index storage through on-demand recomputation~\cite{wang2025leann,seemakhupt2024edgerag}; \rev{LLM-serving caches store responses or KV states per application~\cite{bang2023gptcache,jin2026ragcache}. Concurrent visions have origins expose chunks behind a DNS-like semantic resolver~\cite{bilal2025ainative} or distribute KV caches like CDN content~\cite{cheng2024kdn,ray2026kvinternet}, and a major CDN now builds opt-in per-site AI indexes~\cite{cloudflareaiindex}. SemDN instead delivers content-level evidence ranked across origins, for clients as well as origins:} it joins semantic selection with an origin-authorized hierarchical delivery and freshness path, sharing acquisition and normalized content across tenants while exposing semantic demand to placement and refresh decisions.

This paper makes three contributions.
\textbf{(1)} We characterize today's LLM web-retrieval path and distill four structural mismatches between that workload and the search-engine/CDN stack, including its redundant, failure-prone preprocessing (\S\ref{sec:obs}).
\textbf{(2)} We propose Semantics Delivery Network (SemDN), a new edge retrieval substrate, and contrast its data path with today's (\S\ref{sec:scdn}, Fig.~\ref{fig:arch}).
\textbf{(3)} We organize its research agenda around shareable retrieval state, hierarchical caching and refresh under coverage risk, and two-sided, verifiable deployment (\S\ref{sec:design}), and report preliminary evidence (\S\ref{sec:results}): page-first retrieval processes two to three orders of magnitude more content than agents consume, agent queries exhibit task-local chunk reuse invisible to URL caches, and chunk delivery matches or beats page retrieval at far smaller budgets.

\section{Background}
\label{sec:bg}

\subsection{Content Delivery Networks}

A CDN caches origin content at edge points of presence (PoPs), keyed by URL and expired by a time-to-live (TTL)~\cite{ali2025cdn}. Its value is two-sided in effect but one-sided in billing: clients enjoy lower latency, while content owners, who pay the CDN, offload origin traffic, absorb flash crowds, and gain DoS protection.
However, existing CDNs treat content as opaque bytes, placing and evicting objects by popularity and recency instead of semantic relationships; although CDNs now run more origin logic and defense~\cite{cloudflare-worker-example,lin2025preacher,xiao24snatch,xiao25tocs}, placement remains URL-based and semantics-agnostic.

\subsection{Retrieval for LLMs}

RAG augments an LLM by retrieving external passages from proprietary sources or the live web, and placing them in the model's context~\cite{lewis2020rag}. A corpus is segmented into \emph{chunks} (typically a few hundred tokens~\cite{zhang2025sage,qu2025semantic}), each mapped to a dense vector by an \emph{encoder}, e.g., E5~\cite{wang2022e5} or BGE~\cite{chen2024bge}. A query is embedded by (typically) the same encoder, and an approximate nearest-neighbor search (ANNS) system returns the top-$k$ chunks using a graph index such as HNSW~\cite{malkov2018efficient,liu2025dhnsw}, a clustering-based index~\cite{jegou2011ivfpq,baranchuk2018imi}, or low-bit representations~\cite{ge2013opq,zhan2021jpq,zhan2022repconc,hua2026matryoshka,hua2026spruce}.

\emph{Agentic search} 
interleaves reasoning and retrieval over multiple turns: the agent identifies missing information, queries, incorporates returned passages, and repeats, turning one question into a task-coherent sequence~\cite{jin2025searchr1,asai2024selfrag,wu2025webwalker}. Yet similar passages may carry nothing useful and can mislead reasoning~\cite{cuconasu2024noise}, so the best retriever may be trained on downstream correctness and coupled with the agent~\cite{liu2026agenticr,shi2024replug}.

\subsection{How LLMs Retrieve From Web Today}
\label{sec:bg:today}

A common production path for LLM web retrieval is a two-stage pipeline (Fig.~\ref{fig:arch}a). A search engine results page (SERP) API (e.g., Serper~\cite{serper}, Brave~\cite{bravesearchapi}, Exa~\cite{exa}) maps a query to ranked URLs and snippets but does not return content. A separate reader (e.g., Firecrawl~\cite{firecrawl2026}, \rev{Crawl4AI~\cite{crawl4ai2026},} Jina~\cite{jinareader,wang2025readerlm}, Tavily~\cite{tavily2026}) fetches those pages, strips boilerplate, and returns clean text. The application then chunks, embeds, ranks, reasons, and may issue another query~\cite{liu2026agenticr,jin2025searchr1,song2025r1searcher}. Without cross-provider reuse or origin-integrated freshness, each application assembles its own pipeline or adopts a provider's.

\section{Observations}
\label{sec:obs}

We distill the gap between this workload and its infrastructure into four observations.

\underline{\emph{O1: Objective mismatch—similarity is not utility.}} 
Mainstream discovery services~\cite{serper,bravesearchapi} return results ranked by generic relevance (such as lexical similarity), while off-the-shelf encoders~\cite{reimers2019sentence} optimize query--passage \emph{similarity}; agents instead need passages that are \emph{useful} for producing a correct answer.
The two differ: similarity-trained retrievers underperform utility-aware ones~\cite{liu2026agenticr,zhao2026r3ag}, and highly similar passages can even derail reasoning~\cite{cuconasu2024noise}. Search-R1 adapts a search policy to a fixed retriever with answer-level RL; Agentic-R also trains the retriever on answer correctness~\cite{jin2025searchr1,liu2026agenticr}. This trend motivates tenant-specific utility ranking rather than treating shared similarity as the final objective. Content acquisition and canonical candidates can be shared; a tenant's utility function and outcome labels need not be.
\rev{In our probes, similarity-ranked page-first retrieval leaves useful chunks unreached and is matched or outperformed by chunk-first retrieval (\S\ref{sec:results:e4}); whether utility-customized pipelines widen this gap is future work.}

\underline{\emph{O2: Granularity mismatch—pages versus chunks.}} The web is authored, crawled, and delivered at the granularity of pages and their resource-heavy objects, whereas LLMs consume shorter passages or chunks~\cite{qu2025semantic,zhao2024metachunking,almaliki2026caas}.
An HTML-only fetch still transfers the full document; browser-rendering agents may additionally retrieve its subresources—the median desktop page is now $\approx$3\,MB across roughly 70--77 requests~\cite{pageweight}, while the passage a model wants is typically a few kilobytes.
Many production agents even render complete pages and operate over \emph{screenshots} or other browser-level representations~\cite{openai-agent,claude-visual,google-visual,koh24visualwebarena,openai-visual,wang2026pixelrag}.

Chunking is also a \emph{budget} decision: stuffing whole pages into the model's context inflates inference cost and can degrade answer quality~\cite{liu2024lostmiddle}. How content is chunked is itself improvable, from fixed-size windows and neural text segmentation~\cite{lukasik2020crossseg,qu2025semantic} to LLM-driven chunkers that split at logical boundaries or into self-contained propositions~\cite{duarte2024lumberchunker,zhao2024metachunking,chen2024densex,chen2025index}, and the best granularity varies per query~\cite{zhong2025mog,yeo2026universalrag}.
Existing mitigations optimize this work on the client~\cite{bohra2025weblists,tan2025htmlrag,chen2025index,liu2026dripper} \rev{or convert whole pages to Markdown at the CDN~\cite{cloudflaremarkdown}}; we ask whether a shared edge substrate can instead perform it once, well, and amortize it (\S\ref{sec:scdn}).

\underline{\emph{O3: Locality and policy mismatch—URL versus semantics.}}\\ CDN caching keys on URL and recency; LLM retrieval keys on semantics and can exhibit locality that the current stack cannot see. \rev{AI scrapers and search agents already degrade URL caches: even a search agent draws 70--100\% unseen URLs within a task~\cite{zhang2025aiwebcache}, yet in our traces 32--55\% of a new query's chunks recur within the task (\S\ref{sec:results:e2}). Reuse invisible to URLs reappears at chunk granularity.}
Query streams in RAG workloads are bursty and task-clustered~\cite{wu2025webwalker,lee2026agenttrace,deng2026lever}: within one task and across similar tasks, retrieved chunks recur across turns. Separating discovery from URL-object delivery prevents the delivery layer from directly exploiting this semantic locality for cross-client placement and eviction.

\underline{\emph{O4: Redundant, failure-prone semantic preprocessing.}} Turning a URL into model-ready chunks is a heavy, multi-stage pipeline that requires reaching the page (often past bot walls~\cite{ousat2026brokengates}, JavaScript shells, and timeouts), extracting content from a noisy web, stripping boilerplate, segmenting, and embedding~\cite{steiner2026web,foley2026wcxb,chen2025index,wang2025readerlm,tan2025htmlrag}.
If each application assembles this pipeline itself (\S\ref{sec:bg:today}), popular pages may be processed independently by many operators.
The work is both substantial and unreliable: \rev{across 1,430 live SERP result URLs, a cloud renderer (Cloudflare Browser Rendering~\cite{cloudflarebrowserrendering}) obtains usable clean text from at most 75\%}, after which the text must still be de-boilerplated and embedded---work repeated across tenants within a single freshness window while origins absorb the resulting fetch storms~\cite{fastly}.

\section{Semantics Delivery Network (SemDN)}
\label{sec:scdn}

\subsection{Architecture}

\rev{SemDN serves two kinds of customers. Origins enroll, as with a CDN, for offload, freshness management, and protection from machine traffic. Clients, or \emph{tenants}, choose how enrolled content is chunked, encoded, and ranked, within the transformations origins permit.}
In the data path (Fig.~\ref{fig:arch}b), a client (e.g., a RAG application or an agent) sends a \emph{query}, not a URL, to its nearest edge.
We assume a mature SemDN already maintains a large origin-authorized corpus, populated by enrolled origins and prior acquisition, rather than building it per query. The edge runs three core functions: \textsf{encode} (map queries and content candidates to vectors), \textsf{search} (over canonical or registered model-specific indexes), and \textsf{cache} (store and evict normalized blocks, materialized chunks, and index state). Borrowing the hierarchy that makes CDNs scalable, an edge PoP holds hot content and index state, regional parents hold larger, colder shards, and a global tier maintains the distributed catalog; resource selection can avoid searching every shard~\cite{aly2013taily}.

This hierarchy changes the meaning of a cache miss. A URL cache\rev{, like an information-centric network that routes on exact content names~\cite{jacobson2009ccn},} has a binary exact-match miss; \rev{semantic search replaces names with ranked queries and always returns top-$k$ results}, even when newer or better evidence lies outside its index. We define a \emph{semantic coverage miss} to occur when the estimated risk that external discovery would add fresh evidence and change the returned top-$k$ exceeds a policy threshold. The coverage controller could combine retrieval-score distributions and result coherence~\cite{vlachou2024denseqpp,tian2025agenticqpp,tian2026retrievalutility,sinha2026ragqpp}, \rev{judgments of whether the retrieved evidence suffices to answer~\cite{joren2025sufficient,chen2026rinse},} cross-tier agreement, and freshness metadata as imperfect signals of coverage risk. \rev{A judgment reading only the query and evidence, not the tenant's generator, can run once in the network before generation.} Periodically, or on a high-risk query, it invokes an external discovery service (e.g., a SERP API)\rev{, as Corrective RAG does for low-confidence retrieval~\cite{yan2024crag}}; results are filtered against the origin-enrollment registry before SemDN fetches or revalidates absent or stale content. \rev{FreshCache gates per-client semantic-cache reuse on an estimated staleness probability~\cite{mansoor2026freshcache}; SemDN applies a similar risk test to a shared, origin-authorized corpus.} Calibrating this risk under workload and content drift is an open problem.

\subsection{Design Decisions}


\emph{Share normalized content and canonical candidates; specialize ranking per tenant (O1).} Different chunkers and encoders induce different objects and proximity graphs, \rev{and a single vector index cannot express every top-$k$ ranking~\cite{weller2025limits}}. SemDN caches versioned normalized content blocks (e.g., DOM paragraphs, table blocks, image regions, and provenance) from which supported chunking policies compose passages.
A shared canonical index retrieves candidates; a tenant-provided utility reranker or a pre-registered model-specific encoder refines them, with specialized index state materialized only where demand justifies it. Tenant utility functions and outcome labels remain private (\S\ref{sec:design:representation}).

\emph{Serve and cache at chunk granularity (O2).} The edge returns passages, not pages. Normalized blocks remain the stable shared objects; hot chunks composed under supported policies can be materialized and cached at the granularity the workload actually reuses (\S\ref{sec:results:e1},~\S\ref{sec:results:e2}).

\emph{Place and evict in the query stream (O3).} Because SemDN sees the agent query stream, it can place and evict based on observed semantic reuse rather than URL popularity alone. How much query state to share across tenants is open (\S\ref{sec:design}).

\emph{Do shared work once and specialized work once per registered representation (O4).} Acquisition, cleaning, and normalization run once per content version; canonical embedding and indexing are shared across tenants, while registered model-specific state is amortized among tenants that use it. This relieves origins of redundant fetches (\S\ref{sec:results:e2}).

\section{Architectural Research Agenda}
\label{sec:design}

SemDN raises three coupled questions: what content and retrieval state is \textbf{shareable} across heterogeneous tenants; how a hierarchical substrate stays \textbf{manageable} in caching and refreshing under semantic coverage misses; and what interfaces, incentives, and guarantees make it \textbf{deployable}.

\vspace{-0.06in}
\subsection{What Retrieval State Is Shareable?}
\paragraph{Representation.}
\label{sec:design:representation}
What stable \emph{form} should the cached asset take? Beyond the versioned normalized blocks of \S\ref{sec:scdn}, a canonical embedding or a Matryoshka representation, truncatable to smaller dimensions~\cite{kusupati2022matryoshka,rege2023adanns}, enables a shared first-stage candidate index but presumes an agreed encoder, or a bridge to it (\S\ref{sec:design:bridge}).
Even with recompute, the proximity graph is \emph{metric-specific}: an HNSW graph encodes nearest-neighbor relations in one embedding space and does not transfer to another~\cite{malkov2018efficient,wang2025leann}, so a model-specific graph is shared only among tenants registered for the same representation.

\paragraph{Indexing and the storage-compute trade.}
\label{sec:design:index}
SemDN's baseline uses a canonical index for candidate generation; sufficiently popular registered encoders may add model-specific indexes. For either option, storing embeddings versus recomputing them remains flexible. LEANN~\cite{wang2025leann} and MHR~\cite{hua2026matryoshka} can reduce the storage requirement of graph-based indexes through lightweight structures~\cite{jegou2011ivfpq}, at the expense of additional GPU load. Where a real edge should sit on this surface and whether PoPs carry the assumed compute is left open.

\paragraph{\rev{Bridging encoders to the canonical index.}}
\label{sec:design:bridge}
\rev{A canonical index serves a tenant with its own encoder only if the two embedding spaces can be bridged. Compatible training makes a new encoder comparable to old embeddings~\cite{shen2020bct,ramanujan2022fct} but presumes control over how encoders are trained, which bring-your-own tenants deny. Query-side adapters instead project a tenant's query into the indexed space~\cite{vejendla2025drift,chung2026reranking}, but an adapter fit to regress the tenant space onto the canonical one can undo much of the tenant's fine-tuning. Unsupervised translation reaches unrelated backbones without paired data~\cite{jha2025vec2vec}, which also means tenant query embeddings are not opaque to the operator. All of these treat stored vectors as the only artifact. Because SemDN keeps normalized blocks and can recompute, it can instead decouple recall from ranking: the canonical index, possibly alongside an encoder-agnostic lexical index, need only place a tenant's true top-$k$ within its top-$K$ candidates ($k \ll K$), which the registered encoder re-encodes and ranks exactly. Re-encoded vectors of hot chunks are cached per encoder and grow into the model-specific index above where demand justifies it. How candidate recall degrades with encoder drift, and when re-encoding stops paying off, remain open.}

\vspace{-0.06in}
\subsection{How Should a Hierarchical SemDN Cache and Refresh?}
\paragraph{Caching, locality, and coverage.}
This is the richest open area (O3). \rev{Where AI-aware web caches filter training scrapers and search agents alike as low-locality load~\cite{zhang2025aiwebcache}, SemDN serves query-time retrieval, not training scrapes: an agentic task, or an organization's agents, forms a cohort whose reuse follows semantics.} Agent trajectories reveal which chunks are reused together~\cite{lee2026agenttrace}, while cross-tenant demand determines which content merits edge placement rather than a regional-only copy. \rev{Approximate reuse of retrieval results across similar queries~\cite{bergman2025proximity,oh2026aker,guo2018foggycache} could extend hits beyond exact repeats.}

The challenges include (1) \emph{coverage skew} (only what is queried gets optimized, leaving the long tail under-served); (2) \emph{cross-tenant leakage} (reweighting a shared structure from one tenant's trajectories can expose its task structure to others); and (3) \emph{feedback entrenchment} (favoring previously successful chunks risks locking in stale answers). The controller must also distinguish an edge miss, which a regional or global parent can satisfy, from a system-wide semantic coverage miss that warrants external discovery: false confidence hides fresh evidence, \rev{and stale evidence mixed with current evidence degrades answers~\cite{ouyang2025hoh},} while a conservative threshold restores redundant search and origin load.

\paragraph{Freshness.}
SemDN can adapt CDN-style freshness machinery to amortize \emph{acquisition}. For example, a query that forces a re-fetch can refresh related chunks (spatial locality), and content warmed by one tenant's traffic serves the cohort. \rev{Versioned blocks also suit evolving documents, where amendments change few clauses but retrieval must select the right version~\cite{nam2026timelyrag}.}
However, freshness-critical content (news, prices) limits cache lifetimes; \rev{periodic TTL refresh provably leaves stale windows~\cite{li2026evidex}, graph indexes must absorb frequent in-place updates~\cite{liu2026yi},} and cold content may never be warmed by piggybacking. Therefore, we treat the achievable savings as something to estimate rather than assume.

\vspace{-0.06in}
\subsection{What Interfaces, Incentives, and Guarantees Make SemDN Deployable?}
\paragraph{Tenant interface.}
\label{sec:design:transparency}
Today's agents retrieve from the web through SERP and reader APIs (\S\ref{sec:obs}) or managed indexes~\cite{cloudflareaisearch}. They return ranked snippets, cleaned text, or a synthesized answer \rev{with citations, as search engines' grounding APIs do~\cite{googlegrounding2026},} along with a bill, but generally expose limited control over \emph{why} a passage was returned or how retrieval policy affects it. Yet the embedding model and chunking strategy materially change which passages are retrieved and how well the agent answers~\cite{muennighoff2023mteb,caspari2024beyond}, so a fixed, hidden choice can cap quality, and a per-query bill prices a process the client cannot inspect.

SemDN instead lets a client select a supported chunking policy over normalized blocks and either use canonical retrieval with a tenant utility reranker or register a model-specific encoder (\S\ref{sec:design:representation}). It returns selected passages with scores and provenance, making retrieval auditable and billable for what was consumed.
Managed services already offer website ingestion, configurable chunking/models, reranking, citations, semantic query caching, and tenant isolation~\cite{cloudflareaisearch}; \rev{CDN operators add opt-in per-site AI indexes with change feeds~\cite{cloudflareaiindex}. Both are organized around the origin: one index and representation per site, federated by fan-out. SemDN instead ranks chunks across origins (\S\ref{sec:results:e4})} and shares origin-authorized acquisition and normalized content across independent retrieval clients, coordinating placement, freshness, and representation-specific state around client demand. \rev{For example, to compare two vendors' API limits, an agent today fetches and chunks five pages per turn; AI Search answers only from sites that deployed it, one index each; SemDN returns one top-$k$ across both origins in the agent's chosen representation.}

\paragraph{A two-sided service.}
\label{sec:design:cost_model}
\rev{A CDN is a reverse proxy paid by the origin; an enterprise web proxy is a forward proxy for its clients. \textbf{SemDN is both and can monetize two services.}} To clients, including organizations whose agents share locality, it sells \emph{retrieval quality and latency}: instead of fetching full pages and running the whole pipeline, RAG systems receive model-ready chunks, billed per chunk consumed and priced along the representation/accuracy surface (\S\ref{sec:design:representation}).

To content owners, it sells origin offload for machine traffic, analogous to a CDN: SemDN terminates agent fetches at the edge and serves only the relevant chunks rather than full pages, reducing both origin load and delivered bytes while retaining freshness management and DoS defense. An origin contract defines enrollment, permitted transformations, freshness/version bounds, purge and invalidation, access control \rev{(agent traffic is already identifiable at the TLS/HTTP layer~\cite{kang2026whoseagent})}, and provenance; \rev{retrieval configuration is not part of it, being settled between SemDN and each client.} The interface could expose aggregate machine-demand statistics while withholding individual queries. Pricing can also steer architecture: charging registered model-specific representations at their unshared cost while discounting a canonical representation gives tenants an incentive to use shareable state. \rev{Settlement is the main adoption hurdle: neither side pays merely because the other uses SemDN.}

\paragraph{\rev{Verifiability.}}
\label{sec:design:trust}
\rev{Acting for both sides creates conflicts a CDN has never faced: an operator paid by origins may rank paying origins higher, and a shared cache can leak tenants' queries through timing, as in shared prompt caches~\cite{gu2025promptcache}. HTTPS authenticates only the endpoint and byte stream: a CDN already terminates TLS with the origin's delegated credentials~\cite{rfc9345}, and a chunk is byte-identical to nothing the origin published, so a faulty edge can cherry-pick, suppress, or inject passages~\cite{zou2025poisonedrag}. Clients must instead verify the \emph{evidence-construction process}: that blocks come from their origin at a stated version (e.g., via HTTP message signatures~\cite{rfc9421}); that a chunk is composed of those blocks (via Merkle inclusion proofs); that the top-$k$ is the honest result over a committed index, which also exposes pay-to-rank bias~\cite{chen2014authtopk,qiu2026v3db,avramov2026atlas}; and that the index omits no enrolled origin, as transparency logs ensure for certificates~\cite{rfc9162}. Verifiable HNSW search takes 0.8--2\,s per query~\cite{avramov2026atlas}; amortizing audit-on-demand proofs~\cite{qiu2026v3db} across cohorts and tiers is an open networking problem.}

\section{Preliminary Results}
\label{sec:results}

We report preliminary evidence along three axes. Two concern \emph{cost}: the content volume an agent processes to obtain the text it consumes (\S\ref{sec:results:e1}; O2, O4), and the locality and cross-tenant amortization a chunk cache enables where a page cache cannot (\S\ref{sec:results:e2}; O3--O4).
The third concerns answer \emph{quality} (\S\ref{sec:results:e4}; O1--O2).
Throughout, a \emph{chunk} is a passage of $\approx$100--256 tokens and a \emph{page} is its whole source web page; a \emph{reranker} re-scores retrieved candidates. 

\subsection{The Page Tax: Bytes Processed vs.\ Used}
\label{sec:results:e1}

\begin{figure}[tbp]
\centering
\includegraphics[width=0.8\columnwidth]{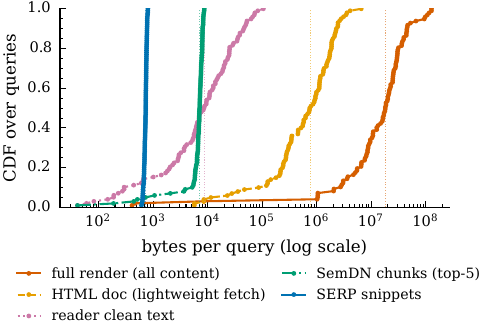}
\vspace{-0.05in}
\caption{CDF of decompressed content bytes per query over the top-5 URLs (live web, 100 queries); values in the text are medians.}
\label{fig:pagetax}
\vspace{-0.1in}
\end{figure}

\begin{figure*}[!t]
\centering
\begin{subfigure}[t]{0.3\textwidth}
  \centering
  \includegraphics[width=\linewidth]{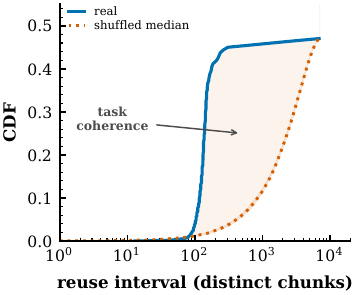}
  \caption{Reuse interval}
  \label{fig:locality-a}
\end{subfigure}\hfill
\begin{subfigure}[t]{0.3\textwidth}
  \centering
  \includegraphics[width=\linewidth]{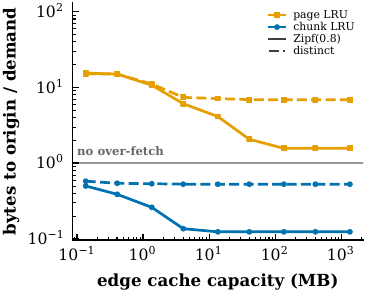}
  \caption{Cache efficiency}
  \label{fig:locality-b}
\end{subfigure}\hfill
\begin{subfigure}[t]{0.3\textwidth}
  \centering
  \includegraphics[width=\linewidth]{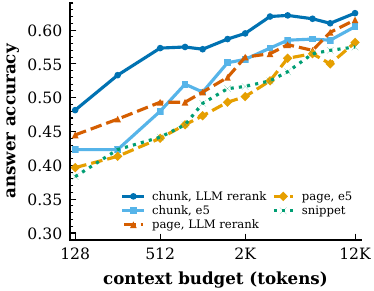}
  \caption{Answer quality}
  \label{fig:isobudget}
\end{subfigure}
\vspace{-0.05in}
\caption{\textbf{Agentic locality (a,b) and answer quality (c).} \textbf{(a)}~Reuse-interval CDF for distinct chunks; shaded band shows 5--95\% over 50 popularity-preserving, task-shuffled controls. \textbf{(b)}~Bytes to origin per demand byte vs.\ edge cache capacity. \textbf{(c)}~Answer quality vs.\ reader context budget (CRAG Task-3, $n{=}600$).
}
\label{fig:locality}
\end{figure*}

We send 100 real queries (50 curated intents across ten categories plus 50 from Natural Questions~\cite{kwiatkowski2019nq}) through the production agent path of Fig.~\ref{fig:arch}a: a SERP API returns the top-5 URLs, which we then fetch and process.
Five distinct byte bands emerge (Fig.~\ref{fig:pagetax}). At the median, the model consumes \textbf{7.1\,KB} of selected chunks, while the lightweight no-JS baseline processes \textbf{761\,KB} of decompressed HTML to obtain them (\textbf{107$\times$}); a full browser render yields \textbf{18.1\,MB} of decompressed content (\textbf{2,532$\times$}, or 5.75\,MB on the wire). Fetching is also brittle: at least 39\% of the 500 sampled URLs lose over half their text without JavaScript and $\approx$19\% are bot-blocked, making fetch-clean-chunk-embed heavy and failure-prone when performed independently by each client (O4).

\subsection{Agentic Locality}
\label{sec:results:e2}

\emph{Observed session locality.} We run Search-R1~\cite{jin2025searchr1} on Qwen2.5-7B~\cite{qwen2025qwen25} over wiki-18~\cite{karpukhin2020dpr,jin2025flashrag} (21M chunks, 3.2M pages) with the fixed e5-base-v2~\cite{wang2022e5} backend to which its search policy was adapted~\cite{jin2025searchr1}. The locality (O3) is large and \emph{task-driven}: within a task, 30.5\% of post-first-turn queries are byte-identical repeats, \rev{consistent with the redundant retrieval independently reported for Search-R1~\cite{sharma2026testtime},} and even after deleting every repeat, 32--55\% of a new query's chunks were already retrieved earlier in the task.
Compared with popularity-preserving, task-shuffled controls, the real stream reuses
chunks after 1--2 orders of magnitude fewer intervening chunks---structure a URL/TTL cache cannot key on (Fig.~\ref{fig:locality-a}).


\emph{Trace-driven cache-byte savings.} We replay the stream against simulated PoP caches. In the evaluated URL-object cache, an entry is a source article sized as the sum of its chunk bytes, and a miss fetches the entire article. This favors the URL-object cache by excluding markup and subresources. A chunk-granularity LRU cache of just \textbf{1.34\,MB} serves \textbf{73.8\%} of chunk requests (87.6\% at 13.4\,MB). The URL-object LRU under page-first delivery reaches a comparable hit ratio only near 100\,MB and fetches \textbf{6.9--15.5$\times$} more bytes than the agent consumes; at equal cache capacity, the chunk cache sends \textbf{13--41$\times$} fewer bytes to origin (Fig.~\ref{fig:locality-b}). The study replicates with the same agent over a realistic CRAG web corpus, where within-task reuse is stronger (63--79\%) and the chunk cache sends 25--70$\times$ fewer origin bytes at equal size.

\emph{Modeled cross-tenant amortization.} The top-10 domains draw 34\% of results across our live set, suggesting potential overlap but not establishing that tenants request the same pages or chunks. We therefore feed the measured per-page and embedding costs (392 chunks/s per A100 GPU with e5-base-v2) into a Zipfian tenant-demand model, which estimates $\approx$6$\times$ less repeated fetch-clean-chunk-embed work at 50 tenants. Embedding and index work amortize only among tenants sharing a representation; a private encoder stays at $1\times$, though acquisition and normalized blocks remain shareable---the incentive for encoder-aware pricing (\S\ref{sec:design:cost_model}).

\subsection{Answer Quality}
\label{sec:results:e4}

On a fixed corpus (wiki-18) with five QA sets (HotpotQA~\cite{yang2018hotpotqa}, 2WikiMultiHopQA~\cite{ho2020twowiki}, MuSiQue~\cite{trivedi2022musique}, Bamboogle~\cite{press2023bamboogle}, and NQ~\cite{kwiatkowski2019nq}), \rev{we compare SemDN's chunks with a page-first client} that downloads the top pages and passes the reader their best-scoring chunks. \rev{Both share the ranker (Search-R1's e5, alone or followed by Qwen3-Reranker-4B~\cite{zhang2025qwen3emb}) and the Qwen2.5-7B reader, and a Llama-3.1-70B judge~\cite{grattafiori2024llama3} grades the generated answers with CRAG's metric~\cite{yang2024crag}. Chunks match or beat the client even when it downloads ten pages (accuracy 0.456 vs.\ 0.427 with e5, 0.500 vs.\ 0.493 with the reranker; with a Qwen3-32B reader~\cite{qwen2025qwen3}, 0.497 vs.\ 0.486 and 0.562 vs.\ 0.547), which moves $\approx$1,200$\times$ more bytes (3.8\,MB of HTML vs.\ 3.2\,KB, measured on 250 questions), or $\approx$3,000$\times$ when the reranker reads 30 candidate pages.}
\rev{On 300 time-stable HotpotQA questions, 16--29\% of SemDN's global top-5 chunks lie on pages that a whole-page ranker does not select, even with 20 downloaded pages, which page-first designs would miss~\cite{bilal2025ainative}.}



On the noisier, web-derived CRAG benchmark~\cite{yang2024crag}\rev{, with the same judge,} we answer each question once with a Qwen3-32B reader~\cite{qwen2025qwen3} and vary only how the retrieved content is packed into the reader's context.
\rev{With the same LLM reranker (Qwen3-Reranker-4B~\cite{zhang2025qwen3emb}) scoring chunks and whole pages,} chunks beat pages by 3.5--8.2 accuracy points from 256 to 6,144 tokens and beat SERP snippets at every evaluated token budget (Fig.~\ref{fig:isobudget}). \rev{They even beat pages ordered with answer-bearing first by 6--9 points up to 768 tokens: a page spends budget on unrelated text~\cite{liu2024lostmiddle}. The gap narrows above 8K tokens, where the context already holds most of each question's supporting evidence and further content mostly adds only noise.}

\section{Conclusion}
LLM agents consume task-relevant passages, yet today's web stack discovers and delivers URL-addressed objects. We argue that semantic selection should become a first-class network-delivery abstraction, and we propose SemDN, a substrate that joins a search engine's query interface with a CDN's origin-authorized edge delivery \rev{and could be built atop an existing CDN or as a separate network.
It is more than finer-grained caching because semantic delivery changes three things: misses must be estimated rather than observed, clients choose retrieval representations that origins need not see, and only acquisition and normalization, not indexes or caches, are shared by all.}
Our probes show that page-first retrieval processes far more content than agents consume, that agent queries exhibit task-local chunk reuse invisible to URL caches, and that chunk delivery improves quality per context token.
Open challenges in shareable retrieval state, hierarchical caching and freshness under coverage risk, and deployability, including privacy, incentives, and verifiable evidence construction, define a networking research agenda.

\bibliographystyle{ACM-Reference-Format-compact}
\bibliography{refs}

\end{document}